# Accelerating a Cloud-Based Software GNSS Receiver

Kamran Karimi, Aleks G. Pamir, M. Haris Afzal

Rx Networks Inc.

800-1201 W. Pender Street, Vancouver, Canada

{kkarimi, apamir, hafzal}@rxnetworks.com

**Abstract**

In this paper we discuss ways to reduce the execution time of a software Global Navigation Satellite System (GNSS) receiver that is meant for offline operation in a cloud environment. Client devices record satellite signals they receive, and send them to the cloud, to be processed by this software. The goal of this project is for each client request to be processed as fast as possible, but also to increase total system throughput by making sure as many requests as possible are processed within a unit of time. The characteristics of our application provided both opportunities and challenges for increasing performance. We describe the speedups we obtained by enabling the software to exploit multi-core CPUs and GPGPUs. We mention which techniques worked for us and which did not. To increase throughput, we describe how we control the resources allocated to each invocation of the software to process a client request, such that multiple copies of the application can run at the same time. We use the notion of effective running time to measure the system's throughput when running multiple instances at the same time, and show how we can determine when the system's computing resources have been saturated.

**1. Introduction**

Implementing a GNSS receiver completely in software has received attention due to the flexibility it provides [3]. Adding features, configuration changes, and defect fixing is done easier with a software receiver than a hardware one. The downside is that a software receiver usually processes signals slower than a hardware receiver because it tends to emulate the hardware, which may not be the most efficient way of designing the GNSS in software. This makes performance an important aspect of the design and implementation of a software receiver.

Many operations in a GNSS receiver including but not limited to signal acquisition and tracking are inherently independent of each other and are run in parallel when a standard receiver is implemented in hardware [10]. A software receiver can exploit this same parallel execution possibility and benefit from multi-core CPUs and GPGPUs. For this reason this paper concentrates on parallelizing the execution using CPUs and GPUs. These two models have very different characteristics, which greatly affect the results.

Another possible requirement for a software receiver is the ability to process data in real-time. This requirement is evidently related to the performance aspect, as real time operation implies processing data at least as fast as they are received. [1, 6] are examples of efforts to utilize modern parallel processing hardware to implement real time GNSS software. In this paper we focus on a software receiver which is meant for offline operation, in a cloud environment. However, we need to process many requests which must still be processed in a reasonable amount of time. For this reason, high performance is one of the main requirements of this project. Offline processing provides opportunities that we have exploited for performance increase.

The target application is intended to be run in a cloud environment, where data, recorded on many clients, are received and processed. The results are then returned to the client to either directly provide the position estimates or assist with satellite acquisition. While real-time processing is not a strict requirement, reducing response time and increasing total throughput are very important. Not only each client must wait as little as possible to receive a response (low response time), but the system as a whole must make sure that as many requests as possible are processed per unit of time (high total throughput). In order to achieve these goals, the application must be able to fully utilize the available hardware.

Since many instances of the application may be running at the same time, care should be taken to make sure all computational resources are used effectively and without conflicts. For example, starting many instances of the application with each of them running on all cores on a CPU with low first or second level cache may cause cache conflicts, where each thread would invalidate cache data from other threads.

Cloud server instances running this application may be with or without GPUs. Servers are started as client requests increase. Each server then may run many instances of the application to process the requests. The application is passed a number of arguments that determine which resources are to be employed by it to process a request.

When discussing GPUs, we focus on NVidia products because they were available to us for development, testing, and deployment. We chose the CUDA programming toolset because it appears to provide better performance than other GPU programming toolsets [8]. We employed CUDA 5.0, the latest version available at the time [14]. The paper's descriptions and limitations may or may not apply to other GPU products or programming languages.

The rest of the paper is organized as follows. In Section 2 we provide a brief introduction to the peculiarities and differences between GPU and CPU computing. Section 3 describes the application we targeted for acceleration. Sections 4 and 5 provide details on our efforts to make the application run faster on CPUs and GPUs, respectively. In Section 6 we summarize the paper and provide a list of future work.

## 2. CPU and GPU background

Methods of increasing software performance on a CPU include algorithmic optimizations, tuning the code manually or automatically by the compiler to minimize wasted cycles, using vectorization to speed up mathematical processing on each core, and using multiple cores to run portions of the code in parallel. Parallelization has a big potential for performance improvement because the current focus of CPU manufacturers is on increasing the number of cores instead of adding complexity and speed to each core. Effective utilization of the available CPU cores depends to a large degree on the design of the application. In a multi-threaded application each core can run different, independent sections of code. The less synchronization and data exchange are needed among the threads, the better the resulting speedup [2].

A newer approach to software performance is the use of GPGPUs. Here hundreds or thousands of smaller and slower GPU cores are used to perform the same operation on different data. This is usually called the Single Instruction, Multiple Data (SIMD) paradigm [4]. Since GPUs do not share memory with the rest of the system, the challenge with GPGPU programming is minimizing the effects of data transfer to and from the graphics card. Data copying overhead can potentially undo any gains from running a particular code in parallel on a GPU.

Recent GPU cores are capable of running nearly any application. For example, Kepler cards from NVidia [4] support recursion, removing some of the last remaining limits compared to CPUs. However, running an application on a single GPU core will result in a slow execution due to the core's limited performance. On the other hand, GPUs have been successfully used to speed up mathematical operations that process large amounts of data. For example, performing a pair-wise vector multiplication or a dot product on large input arrays will potentially be sped up when using hundreds or thousands of GPU cores, where each core performs a few simple operations, on a small range of the data, at the same time. Using this massive parallelism to speed up the execution of specific portions of the code is the key to obtaining good speed up with a GPU.

The system we used for tests reported in this paper had an Intel Xeon E5507 CPU with 4 cores (not hyper-threaded), running at 2.27 GHz, and 8 GB of RAM. It could boot into 64-bit Windows 7 and Linux. The GPU used was a NVIDIA GeForce GTX 550 Ti with 192 cores and 1GB of memory.

**3. Rx Network's software navigation solution**

Figure 1 depicts the overall architecture of software GNSS receiver's signal processing modules. Depending on the current status of the receiver, either acquisition or tracking channels are active for a satellite at an instance. Normally, acquisition channels are triggered first, producing the coarse signal detection parameters, namely signal Doppler and code phase. These parameters are fine tuned at a later stage with the help of tracking channels. There exists an acquisition and a tracking channel for each satellite and they all are independent of each other, hence they are good candidates for parallelism.

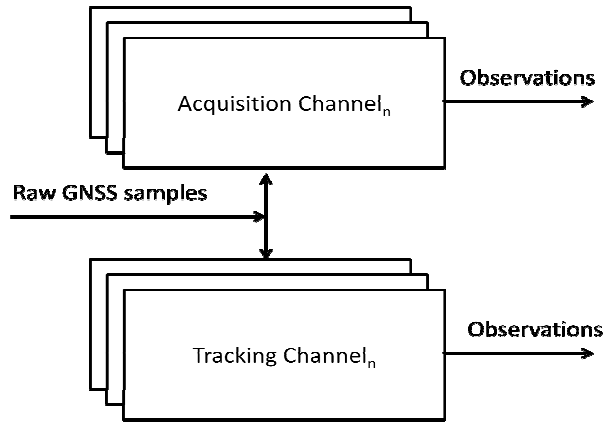

Figure 1. Acquisition and tracking channels of software GNSS receiver.

Figure 2 shows the internal architecture of an individual acquisition channel. The greyed modules are all candidates for parallelism internally to each acquisition module. Both carrier and code Numerically Controlled Oscillators (NCOs) generate local replicas of carrier and code signals for a particular satellite. An iterative loop is utilized for generating these replicas and the number of iterations depends on raw GNSS signal sampling frequency as well as the duration of data utilized for the acquisition operation. For example, if the sampling frequency is 8.184 MHz and 10 ms worth of data is utilized for acquisition, both carrier and code replica loops need to be 81840 iterations long. The mixing stage right after carrier NCO is mathematically a point by point multiplication. Considering the aforementioned example, a total of 81840 independent multiplications need to be performed at this stage, making it an ideal candidate for parallelism. There exist two Fast Fourier Transform (FFT) operations, one after the mixing stage and the other after code replica generation. FFT operations has some parallelism which can be leveraged by developers. After FFT operations, another mixing stage is encountered which is similar to the first stage and a good candidate for parallelism. Both FFT operations and the mixing stage constitute the frequency domain correlation operation utilized by the acquisition channels. In order to utilize the correlation results for signal detection, an Inverse FFT (IFFT) operation is required to represent the results in time domain. Similar to FFT, IFFT also presents some opportunities for parallelism.

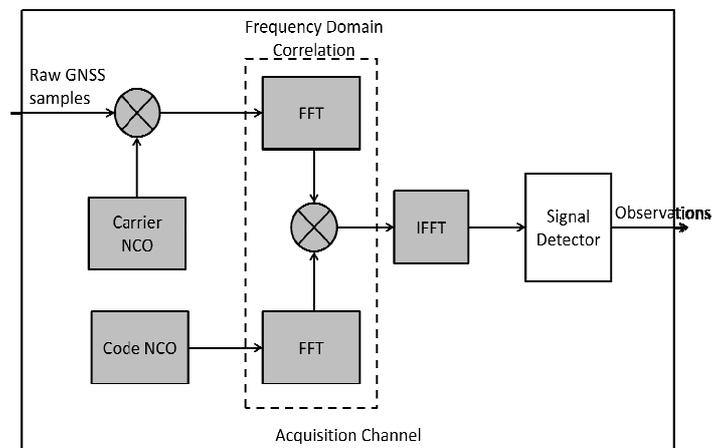

Figure 2. Anatomy of acquisition channel.

Figure 3 shows the detailed architecture of+ the tracking loops utilized for this software GNSS receiver. There are some commonalities among acquisition and tracking modules in the early stages of signal processing. These include carrier and code NCOs as well as the first mixing stage. All of these modules are good candidates for parallelism as is explained earlier. The nature of tracking loops does not allow for the use of frequency domain correlation operation and is therefore computationally more expensive than acquisition. The mixing stage of time domain correlation is the main culprit for this computational burden. Considering the same example of 81840 samples per tracking operation, one has to perform 81840 point by point multiplications and 81840 sample shift operations 81840 times over. So for one tracking loop iteration, a total of $6.7 \times 10^9$ independent multiplication operations are needed, making the tracking channel an ideal candidate for parallelism.

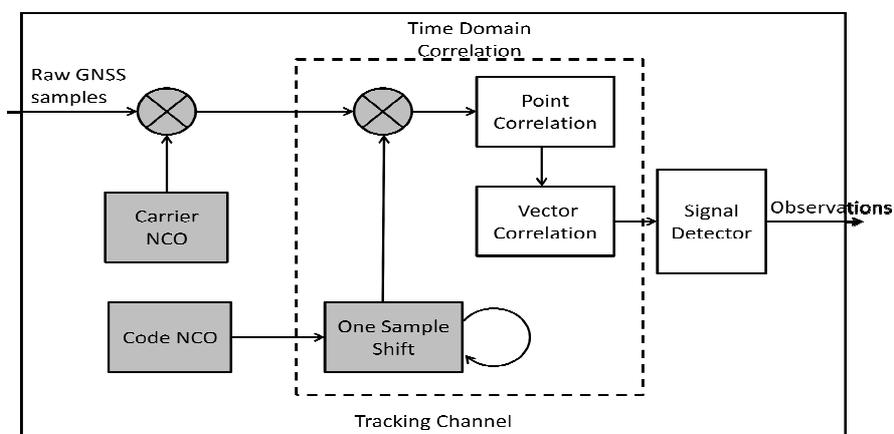

Figure 3. Anatomy of tracking channel.

For various reasons it was important for us to keep the architecture of the code as intact as possible. This requirement had an important effect on our design and implementation decisions, usually limiting our choices, especially with GPUs.

We started with the aim of increasing the performance of the target application, and then ensured high throughput on cloud servers when running multiple instances of this application. The software had already been designed with CPU multi-threading in mind. Native Windows and Linux (pThread) threads were used to run different sections of the code in parallel. Fast Fourier Transform (FFT) functions from Intel Integrated Performance Primitive (IPP) [12] libraries were utilised to perform FFT operations using CPU vectorization [13]. There has been an effort to support GPUs to perform part of the computations under certain conditions, including the presence of strong signals [9].

The CPU threads mimic the parallelism existing in a hardware receiver. Different hardware channels can process input data independently of each other, and this idea closely matches the software design. Our GPU implementation, however, differs from this idea because we cannot run the whole code on a number of GPU cores. Instead, we use a GPU to perform certain mathematical operations in parallel. Each CPU thread can run certain computationally demanding operations on a GPU instead of a CPU.

Doing so utilizes a GPU's parallelism, and at the same time reduces the load on the CPU. Because we did not want to change the application's architecture, GPU code basically mimics IPP code used on the CPU.

**4. CPU Performance**

Our efforts were focused on better utilization of CPU cores, as well as transparent support of GPUs, if available. Algorithmic changes such as block processing resulted in considerable decrease in running time. Beyond that, we replaced most of the loops (iterative schemes) in the code with equivalent IPP functions to benefit from the CPU's vectorization abilities. We noticed that with a small number of iterations, calling an IPP function may take longer than the original loop, so we have a minimum number of iterations, below which normal loops are used, and above which IPP functions are called.

In addition to the native threads already implemented in the code, we added threading mechanisms using OpenMP [2], which has more overhead than native threads, but results in much simpler code. The application uses either native threads or OpenMP threads.

In order to hide OpenMP's extra overhead, parallel loops must be as long-running as possible. If this is not the case, then using OpenMP to parallelize small loops may actually result in a slow-down. The main processing loops of the application looked as in Listing 1 below. In a first attempt, OpenMP threads were used to parallelize the execution of the two for-loops. We noticed a significantly slower execution for OpenMP compared to native threads because the threads were being started and stopped between the for-loops.

```cpp
bool doneProcessing = false;
while( !doneProcessing )
{
    doneProcessing = true;
    #pragma omp parallel for num_threads((numThreads)
    for(int i = 0; i < numTasks; i++)
        task1[i]->Execute();

    // next loop must execute after the first one

    #pragma omp parallel for num_threads((numThreads)
    for(int i = 0; i < numTasks; i++) {
        task2[i]->Execute();
        if(ProcessingComplete(i))
            doneProcessing = false;
    }
}
```
Listing 1. Original code to parallelize the main computing loops of the application using OpenMP

To avoid thread management overheads we kept the threads running outside of the for-loops, and used barriers to make sure the for-loops were executed sequentially. This had a positive impact on code performance. Another modification with high impact was using dynamic scheduling for the OpenMP threads, which allows each thread to be assigned loop iterations as they become available. With the default static scheduling, some threads may run out of work while others are busy with processing loop iterations. We also noticed that assigning high priority to threads reduces the running time because it reduces the number of context switches. Listing 2 below shows the resulting code.

```
volatile bool doneProcessing = false;
#pragma omp parallel num_threads((numThreads)
{
    setThreadPriority(HIGH_PRIORITY);
    while( !doneProcessing )
    {
        #pragma omp barrier
        doneProcessing = true;
        #pragma omp for schedule(dynamic)
        for(int i = 0; i < numTasks; i++)
            task1[i]->Execute();
        #pragma omp barrier
        // next loop must execute after the first one
        #pragma omp for schedule(dynamic)
        for(int i = 0; i < numTasks; i++)
            task2[i]->execut();
            if(ProcessingComplete(i))
                doneProcessing = false;
        }
        #pragma omp barrier
    }
}
```

Listing 2. Improved OpenMP parallelization.

The un-optimized code would create as many threads as channels, so there was a one to one relationship between them. We added the ability to limit the number of CPU threads running at the same time. This was done in consideration for the requirement of having high throughput on the server. Running an application on *N* cores which are not fully utilized and/or in a sustained manner may allow the execution to finish sooner than when running on $M < N$ cores with better utilization. However, we may achieve better throughput when using less cores, because we can run multiple instances of the application at the same time. The result is that multiple instances of the aplication use the CPU fully. Of course each instance will run slower than if it was running on all cores, but throughput will increase.

As a baseline, we ran the original and optimized versions of the application with different number of threads using native and OpenMP threads. The running times, in seconds, appear in Figure 4. Numbers represent the average of 10 runs. We include tests with 12 threads because the particular input file we processed had 12 channels.

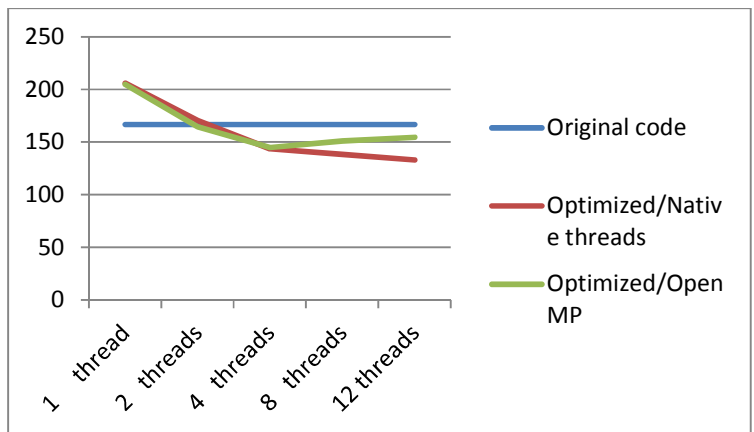

Figure 4. Windows running times, in seconds, with different number of cores

The above Figure shows that, as expected, performance with low number of threads is similar between the two threading mechanisms. For our application, native threads result in better performance than OpenMP as the number of threads increases. This is due to the extra thread management overhead of OpenMP. In our tests we noticed that when running a single instance with 12 threads, using native threads would result in a CPU utilization that varied between 75% and 90%. With OpenMP, CPU utilization was sustained very close to 100%. This indicates that OpenMP threads are more resource consuming, and explains the performance degradation when many of them are running at the same time.

As mentioned before, in the production environment multiple instances of the application will be run simultaneously. We define the *effective running time* be the total running time of *N* instances of the application, divided by *N*, and use it to determine a configuration with the best throughput when running multiple instances of the application at the same time.

To provide a baseline for measuring performance improvements, Figure 5 shows the effective running times when multiple invocations of the original application are running at the same time. Because of memory constraints, we could not run more than 8 CPU instances of the application at the same time.

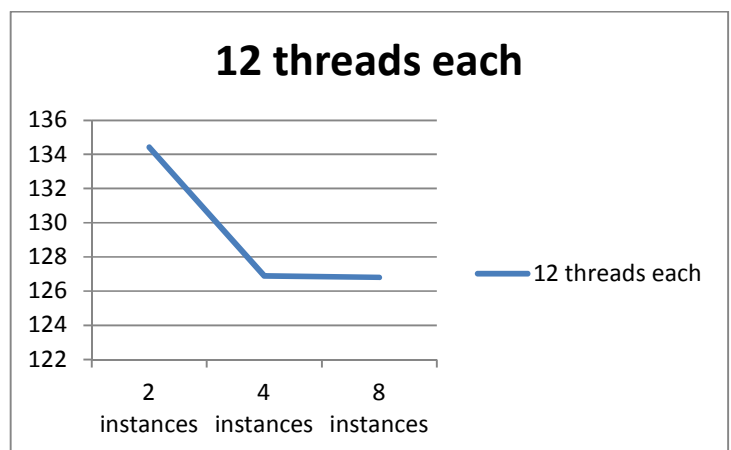

Figure 5. Windows effective running times for the original version.

Figure 6 shows effective running times when different numbers of instances, with varying number of threads, are executed. Native threads were used for these tests because under Windows OpenMP did not provide a clear performance advantage. In our tests, the running times of the concurrent instances were very close, so multiplying the number of instances by the effective running time provides an approximate value for the running time of each instance.

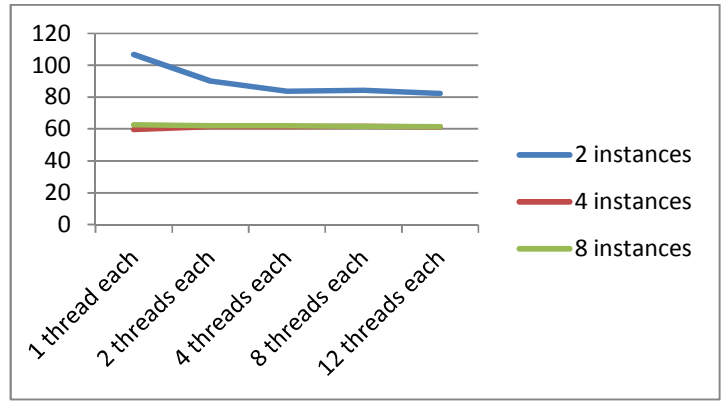

Figure 6. Windows effective running times with native threads.

There is a clear trend in the effective running times. When underutilizing the hardware, effective running times are long because few simultaneous instances are being processed. As the number of simultaneous instances increases, effective running time decreases until it reaches a plateau, which indicates the computer's resource saturation. In this mode, effective running times are close, but it should be noted that with more instances, individual instances take longer to run.

From Figure 6, the optimum configuration seems to be running 4 instances with one thread each, because it provides a low effective running time, and each instance is finished sooner than when more instances are run simultaneously. This observation is easy to justify. On an $N$ core CPU, running $N$ instances in a single thread lessens competition for resources among the threads.

We tried the application with OpenMP threads too, with the results appearing in Figure 7. As can be seen, using many OpenMP threads simultaneously results in a degradation of performance. This is another symptom of OpenMP threads' bigger overhead. Our tests would consistently fail when running 8 instances with 12 threads each because of a lack of resources on the test machine.

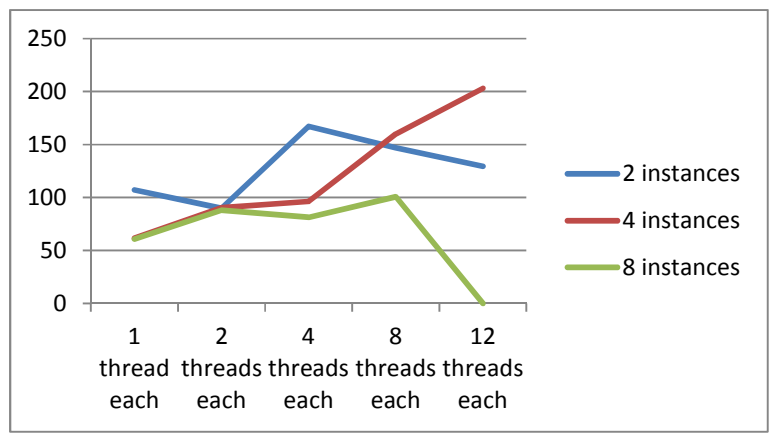

Figure 7. Windows effective running times with OpenMP threads.

Under Windows, OpenMP's performance generally degrades as the number of threads in each instance increases, while the running times with native threads remain relatively stable after a saturation of computing resources.

In a production environment, different instances may be of different importance levels. For example, high priority customers' instances may get priority over others. Figure 7 indicates that if we want some jobs to take shorter to run, we should run fewer simultaneous instances, but with more threads each. In this case system resources will be underutilized.

**4.1 Windows vs. Linux**

Determining the effects of the operating system and compiler on the application's performance is also of interest. Different operating systems' schedulers, memory allocation policies, and I/O driver performance can affect the running times.

The same PC used for the Windows tests was booted into Linux, and we performed the same tests. Figure 8 compares the running times of the original version of the application, hard-coded to run with 12 threads, and the optimized version using pthreads and OpenMP.

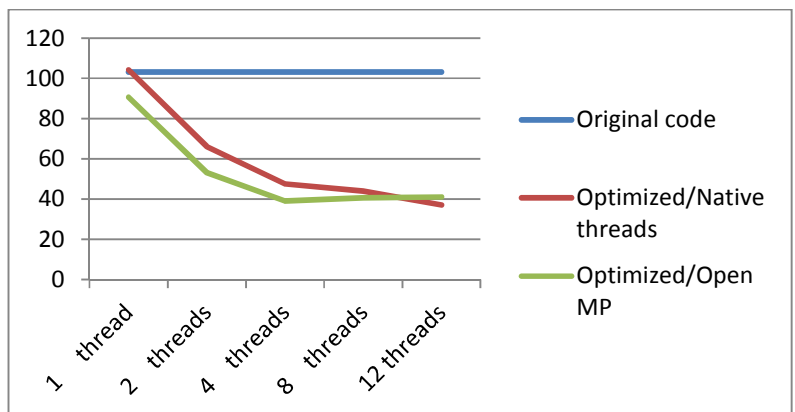

Figure 8. Linux running times of the application with different number of cores

The application performs better under Linux than Windows. Also, OpenMP's performance scales better under Linux.

As a baseline, Figure 9 shows the effective running times when multiple invocations of the original application are running simultaneously under Linux.

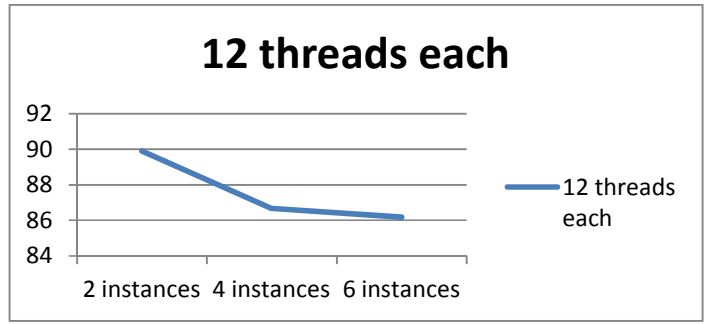

Figure 9. Linux effective running times for the original version.

Because of memory constraints, running more than 6 instances of the original application was not possible under Linux. As can be seen, running multiple instances of the original application reduces the running time, but performance is still less than that of the optimized versions.

Figures 10 and 11 shows the running times for multiple simultaneous invocations of the optimized application, when using native and OpenMP threads respectively.

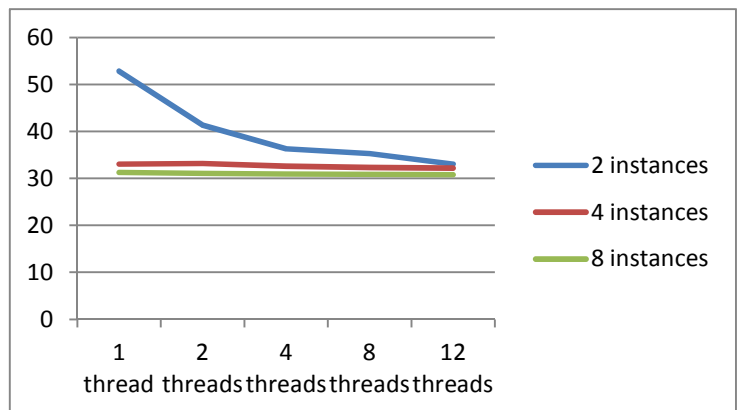

Figure 10. Linux effective running times for the optimized version with native threads.

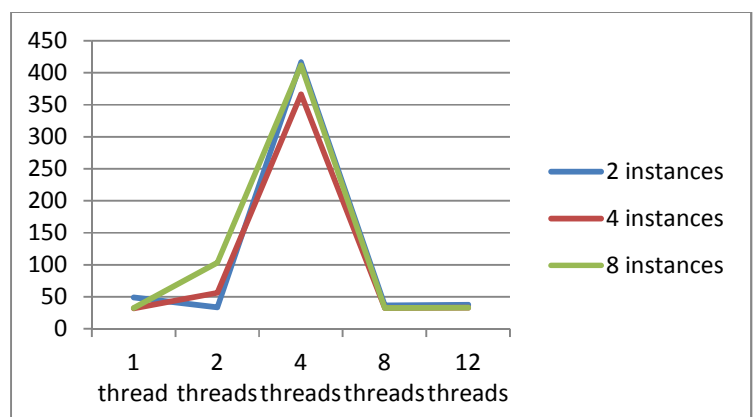

Figure 11. Linux effective running times for the optimized version with OpenMP threads.

With OpenMP under Linux, there is a "resonance" effect happening when each application uses 4 threads. This effect was seen in all of our many trials. Since this effect was not seen under Windows, the cause may be the operating system's cache management and thread scheduling issues. It is interesting to note that when running multiple instances at the same time, OpenMP has no advantage over native threads. We observed that CPU utilization was generally higher when using OpenMP, implying more overhead with OpenMP threads. This observation explains why having many OpenMP threads could reduce performance.

**4.2 Single precision computing**

We performed a number of tests and determined that our application can utilize single precision variables without loss of accuracy. The reason behind this is the limitation on sampling frequency in our application. Due to the limited resolution of the samples coming from the front-end, the receiver is not losing any information by performing signal processing operations in single precision. As a result we decided to use single precision variables in both IPP and GPU computations.

With IPP, we noticed a marked performance improvement when switching from double precision to single precision. This is against the popular belief that double and single precision performances are comparable in modern CPUs, but there are a number of reasons for this improvement. Other than smaller data sizes which fit better in L1 caches, IPP's vectorized instructions can operate on four single precision variables in parallel, vs. 2 double precision variables [15]. On a GPU, using single precision variables allows faster data copying and cache utilization. Additionally, most current GPUs process single precision variables faster than double precision variables.

**5. Performance on GPUs**

GPU programming currently needs considering certain limitations compared with CPU programming. For example, there is usually less memory available on a GPU, and access to a GPU's computing and memory resources involves copying code to execute on the GPU (called a kernel) and data between the CPU and GPU memory. Given our requirement for preserving the multi-threaded architecture of the code, we could not follow the usual route in GPU software design, which is to transfer all data to the GPU, process them using one or more kernel executions, and return the results to CPU-accessible memory, all done within a single CPU thread. Instead, we mirrored the multi-threaded CPU code by identifying the sections of code that were utilizing IPP to process data, and re-implementing the IPP functions with equivalent GPU code.

The CPU code utilizes FFT for frequency domain calculations, and vector multiplications and dot products for time domain computations. Such operations, especially when performed on large sets of data, are good candidates for running on a GPU. We started our work by using libraries such as CUFFT, NPP, CUBLAS [11], and Thrust [7] instead of writing our own custom GPU kernels. The main benefit of doing so was a much shorter development time. With the exception of CUFFT, we noticed a number of problems with these libraries. NPP and Thrust do not support a multi-threaded application in their current versions, so different threads cannot issue GPU operations to be executed in parallel. They do not support CUDA Streams which, as explained below, form the mechanism for identifying GPU code that can be executed in parallel on a GPU. CUFFT does support streams, so we used it for frequency domain computations. CUBLAS supports streams too, but we achieved better performance with our custom kernels. These kernels support streams and perform pair-wise vector multiplication and dot-product for time domain calculations.

Another problem we encountered with GPU computing is high latency in memory copy operations. Our application needed to perform operations on a relatively small amount of data (1 millisecond, or about 8000 single precision variables) at a time. Due to high latency, the time it takes to copy the data to the GPU, and copy the results back, hinders gains from faster computation. We tried to counter this effect by keeping intermediate results on the GPU as much as possible. We also tried to find algorithmic ways of increasing the size of processed data (e.g. batch processing of signals for the entire coherent integration time) to offset the performance losses due to latency [10, 13].

Ironically, the main obstacle to obtaining good speedup with a GPU was the same mechanism for obtaining good speed up on a CPU: multi-threading. Multiple CPU threads active in the application would issue FFT and vector operations independently of each other to the GPU. In a CPU environment the operations can run in parallel in multiple cores, but pre-K20 GPUs have an inherent limitation on the number of threads that can use them at the same time. With most GPUs this limit is one kernel and copy operation at a time. As a result, the many threads of the application have to take turn to copy small amounts of data and run their GPU code, so the GPU becomes a point of serialization. The stream mechanism in a GPU is used to indicate potential parallelism on the GPU [5]. Different instructions running in different streams can potentially run in parallel, but current GPUs ignore this hint from the programmer because of their hardware cannot run multiple streams at the same time.

Starting with Kepler K20 devices, it is possible to run multiple kernels and copy operations in parallel on the GPU, provided they are issued to different streams. Kepler GPUs with Hyper-Q technology [4], can perform up to 32 operations at the same time. This will potentially remedy the problem of the GPU becoming a serialization bottleneck. Lower GPUs do not have this capability, but can still overlap memory copying and kernel executions, so they support a limited form of parallelism, which we exploited by staggering copy and compute operations from different streams in our code and increased GPU performance. However, GPU performance was still not satisfactory. Figure 12 shows the results of profiling a short run of the application. Different threads are performing short bursts of GPU work, which cannot compensate for the overhead of data copy. The bulk of the computation is still done on the CPU.

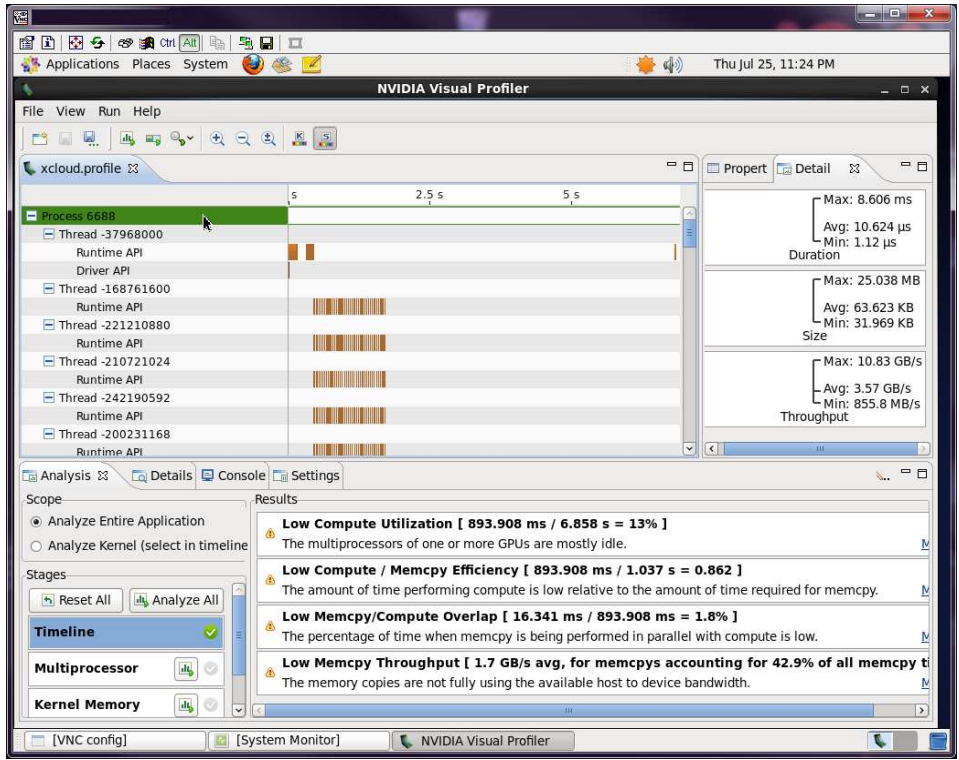

Figure 12. Results of profiling a short run of the application.

We ran the GPU version of the application and recorded the results. For brevity we only report the Linux results. The limited amount of memory in our GPU card limited the number of concurrent executions to two. In one series of runs, we used the GPU to process both the frequency domain (cuFFT) and time domain (custom GPU kernels) sections of the code. The results appear in Figure 13. We then tried the application with the more-efficient frequency domain section being run on the GPU, with the results appearing in Figure 14.

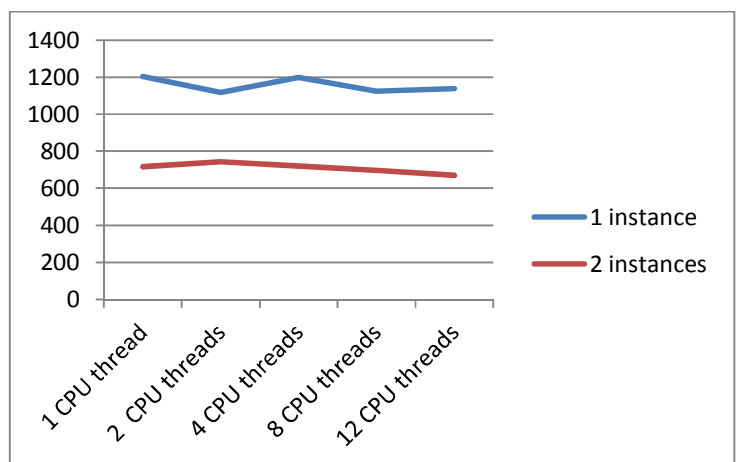

Figure 13. Linux effective running times with both Frequency and time domain calculations on GPU

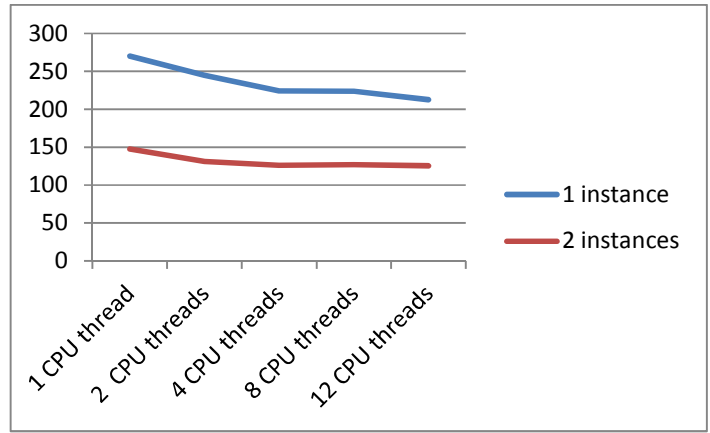
Figure 14. Linux effective running times with Frequency domain calculations only on GPU

In both runs native CPU threads were used. Resulting running times indicate an advantage when running multiple instances of the application on the GPU. GPU utilization values when running one and two instances was close, implying that most of the speedup is due to reduction of the CPU portion of the code's running time, with the GPU still being the point of serialization.

Another observation is that performing time-domain operations on the GPU results in lower performance. The reason is that the speedup of processing 1 ms worth of data on a GPU is not enough to cover the overheads inherent in using a GPU. One solution for improving GPU performance, both for time domain and frequency domain operations is to execute fewer, but bigger operations on the GPU. For example, to perform a single FFT operation on 10 different sets of inputs one can use the pseudo code in Listing 3 (a) or (b). In (a), copying data to and from the card, and performing an FFT operation are done *N* times in independent threads. In (b), the input are coalesced and copied to the GPU once, FFT is performed on all of them using the batch mode (supported by the CUFFT library), and the results are brought back to CPU memory in one operation, thus reducing data transfer latency. The other benefit of doing so is that there is a single invocation of FFT on the GPU, reducing function invocation overhead.

| Each of the *N* independent threads {<br>    Copy data to graphics card<br>    Perform FFT on GPU<br>    Copy the results back<br>} | One single thread {<br>    Place data for all *N* FFTs in contiguous memory<br>    Perform FFT in batch mode.<br>    Copy back all *N* FFTs results to CPU memory<br>} |
|---|---|

Listing 3. (a) Many independent GPU operations. (b) Batch mode operation

The problem with the approach in Listing 3(b) is that implementing it would require a major change in the architecture of the software. Given our requirements, doing so was not possible, so only the algorithm in 3(a) was implemented.

## 6. Summary and future work

In this project we used different optimization techniques to decrease software GNSS application's running time. Working within the constraints of this application, we achieved different degrees of performance gains. Our results on the CPU showed a considerable speedup. We observed that compared to OpenMP, native CPU threads generally provide better performance and a graceful performance drop when compute resources are saturated. However they are harder to implement and tune. Our GPU results beat that of the original code, but lag behind the optimized CPU version.

We ran the same performance tests as above in an Amazon cloud environment. The outcomes, omitted to keep the paper of reasonable length, match what we report in this paper.

Going forward, we are interested in measuring the performance of new-generation GPUs with multiple execution units. Our code is designed with streaming support, and theoretically should benefit from such GPUs. It would also be interesting to see how effective alternative many-core solutions would be for an application such as ours. One such alternative is Intel's Xeon Phi. It contains many (e.g. 32) conventional CPU cores, and thus provides a more familiar environment for high performance computing. A potential problem with such CPU-based solutions comes from Amdahl's law, which is the observation that obtaining appreciable speedup as the number of cores increases is usually a challenge [2], so the effectiveness of such solutions should be verified.

We are also interested in investigating other approaches to increasing GPU performance and utilization by algorithmic improvements, including but not limited to performing low-level signal processing tasks like Doppler removal on GPGPUs.

**Acknowledgment**
We are grateful to Tecterra Inc. for funding the project this research is based on as part of their Industry Investment Program (IND).